\documentclass[amsmath,amssymb,aps,prl,superscriptaddress,twocolumn]{revtex4-1}

\usepackage{graphicx}
\usepackage{dcolumn}
\usepackage{bm}
\usepackage[usenames]{color}
\usepackage{comment}

\usepackage{hyperref} 
\usepackage{bookmark}
\usepackage{mathtools}

\renewcommand{\b}{\beta}
\renewcommand{\d}{\delta}
\newcommand{\e}{\epsilon}
\newcommand{\m}{\mu}
\newcommand{\ph}{\phi}

\newcommand{\s}{\sigma}
\renewcommand{\t}{\tau}

\newcommand{\w}{\omega}

\newcommand{\order}{\mathcal{O}}
\newcommand{\im}{\text{Im}}
\newcommand{\re}{\text{Re}}
\newcommand{\ra}{\rightarrow}
\newcommand{\ua}{\uparrow}
\newcommand{\da}{\downarrow}
\renewcommand{\(}{\left(}
\renewcommand{\)}{\right)}
\newcommand{\<}{\left\langle}
\renewcommand{\>}{\right\rangle}

\newcommand{\lb}{\left|}
\newcommand{\rb}{\right|}

\def\beq#1\eeq {\begin{align}#1\end{align}}
\begin{document}

\title{Linear resistivity and Sachdev-Ye-Kitaev (SYK) spin liquid behaviour in a quantum critical metal with spin-$1/2$ fermions
}

\author{Peter Cha} 
\affiliation{Department of Physics, Cornell University, Ithaca, NY 14853, USA}

\author{Nils Wentzell}
\affiliation{Center for Computational Quantum Physics, The Flatiron Institute, New York, New York, 10010, USA}

\author{Olivier Parcollet}
\affiliation{Center for Computational Quantum Physics, The Flatiron Institute, New York, New York, 10010, USA}
\affiliation{Universit\'e Paris-Saclay, CNRS, CEA, Institut de physique th\'eorique, 91191, Gif-sur-Yvette, France}
\author{Antoine Georges}
\affiliation{Center for Computational Quantum Physics, The Flatiron Institute, New York, New York, 10010, USA}
\affiliation{Coll\`ege de France, 11 place Marcelin Berthelot, 75005 Paris, France}
\affiliation{Centre de Physique Th\'eorique, Ecole Polytechnique, CNRS, 91128 Palaiseau Cedex, France}
\affiliation{Department of Quantum Matter Physics, University of Geneva, 24 Quai Ernest-Ansermet, 1211 Geneva 4, Switzerland}

\author{Eun-Ah Kim}
\affiliation{Department of Physics, Cornell University, Ithaca, NY 14853, USA}

\date{\today}

\begin{abstract}
`Strange metals' with resistivity depending linearly on temperature $T$ down to low-$T$ have been a long-standing puzzle in condensed matter physics. 
Here, we consider a model of itinerant spin-$1/2$ fermions interacting via on-site Hubbard interaction 
and random infinite-ranged spin-spin interaction. 
We show that the quantum critical point associated with the melting of the spin-glass phase by charge fluctuations displays non-Fermi liquid behaviour, with local spin dynamics identical to that of the Sachdev-Ye-Kitaev family of models. 
This extends the quantum spin liquid dynamics previously established in the large-$M$ limit of $SU(M)$ symmetric models, to models with physical $SU(2)$ spin-$1/2$ electrons.  
Remarkably, the quantum critical regime also features a Planckian linear-$T$ resistivity associated with a $T$-linear scattering rate and a frequency 
dependence of the electronic self-energy consistent with the Marginal Fermi Liquid phenomenology. 
\end{abstract}

\maketitle

$T$-linear resistivity is a central enigma of correlated quantum matter.
A universally observed feature of cuprate high T$_c$ superconductors (for a recent review, see e.g.~\cite{proust_taillefer_2019}), it has been reported in several other materials with correlated electrons and has also been the subject of recent investigations in the context of cold atomic gases in optical lattices~\cite{brown2018,xu2019}.
For ``bad metals''~\cite{emery_kivelson_prl_1995,hussey_phil_mag_2004} corresponding to a resistivity larger than the Mott-Ioffe-Regel (MIR) value, i.e. when the nominal mean-free path deduced from the application of a simple Drude formula is smaller than the lattice spacing, this phenomenon can be rationalised using rather general theoretical considerations at high temperatures~\cite{gunnarsson_rmp_2003,palsson_prl_1998,palsson_thesis_2001,deng_badmetals_prl_2013,perepelitsky_hiT_transport_prb_2016,hartnoll_2015,hartnoll2018,cha2019}. 
In contrast, a microscopic understanding remains rather elusive for metals displaying $T$-linear resistivity smaller 
than the MIR value and persisting down to low temperature. 
In pursuit of a theoretical understanding of this puzzle, the idea of the marginal Fermi-liquid (MFL)~\cite{varma1989, littlewood1991} was put forward early on. This approach considers fluctuations with a characteristic energy scale set by temperature itself, leading to a $T$-linear scattering rate $\im\Sigma(\w=0, T)\propto T$. 
This phenomenology lacks a microscopic model in which this is realized, however. 
One strategy towards a microscopic theory has been to investigate the role of quantum critical fluctuations leading to non-Fermi liquid (NFL) behaviour~\cite{marel2003, gegenwart2008}.
However, field theoretic approaches for various itinerant fermion quantum critical points (QCP) typically predict a different power law ~\cite{hertz1976, moriya1985, millis1993} and sign-problem free quantum Monte Carlo found little temperature dependence in the scattering rate~\cite{Lederer4905}.
Hence to the best of our knowledge, microscopic studies of QCP involving itinerant spin-$1/2$ fermions have yet to find a $T$-linear scattering rate.

Another elusive state whose pursuit was motivated by cuprate phenomenology is the quantum spin liquid (QSL)~\cite{anderson1973}.
In efforts to establish a QSL ground state in a microscopic model, Sachdev and Ye (SY)\cite{sy1993} studied a spin model with 
quenched random interactions on a fully connected lattice. 
Remarkably the model has an exactly solvable limit when one 
extends the spin symmetry group to $SU(M)$ and takes the $M\rightarrow\infty$ limit. 
An exciting finding of Ref.~\cite{sy1993} in this solvable limit was a QSL ground state with slowly decaying local spin-spin correlations in the long-time limit $\<S(t)\cdot S(0)\>\sim 1/t$, where $t$ is real time.
Doping this model in the spirit of a $t$-$J$ model, Ref.~\cite{parcollet1999} found, again at $M=\infty$, a QCP separating the SY phase from a Fermi liquid (FL) ground-state.
The quantum critical regime was found to retain the QSL correlations of the SY model and, remarkably, to display `bad metal' behaviour with $T$-linear resistivity in spite of a single-particle scattering rate behaving as 
$\sqrt{T}$~\cite{parcollet1999}.
However, a numerical study of the SY model with physical $SU(2)$ spins found a spin-glass (SG) ordered ground-state instead of the QSL ground-state seen in the large-$M$ limit~\cite{grempel1998}. 
The relevance of SY behaviour to physical spin-$1/2$ electrons and to the $T$-linear resistivity problem in real 
materials is therefore a major open question.

In this article, we provide a major step towards answering this question in the positive by (i) considering a model in which the SG phase can be quantum-melted at the QCP and (ii) providing a numerical solution of this model directly for spin-$1/2$ $SU(2)$ fermions. 
We find, remarkably, that the quantum critical regime displays SY spin-liquid correlations and a scattering rate linear in temperature, leading to $T$-linear resistivity down to $T=0$ at the QCP.
Our numerical results are consistent with the MFL phenomenology.

\begin{figure}
\centering
\includegraphics[width=\linewidth]{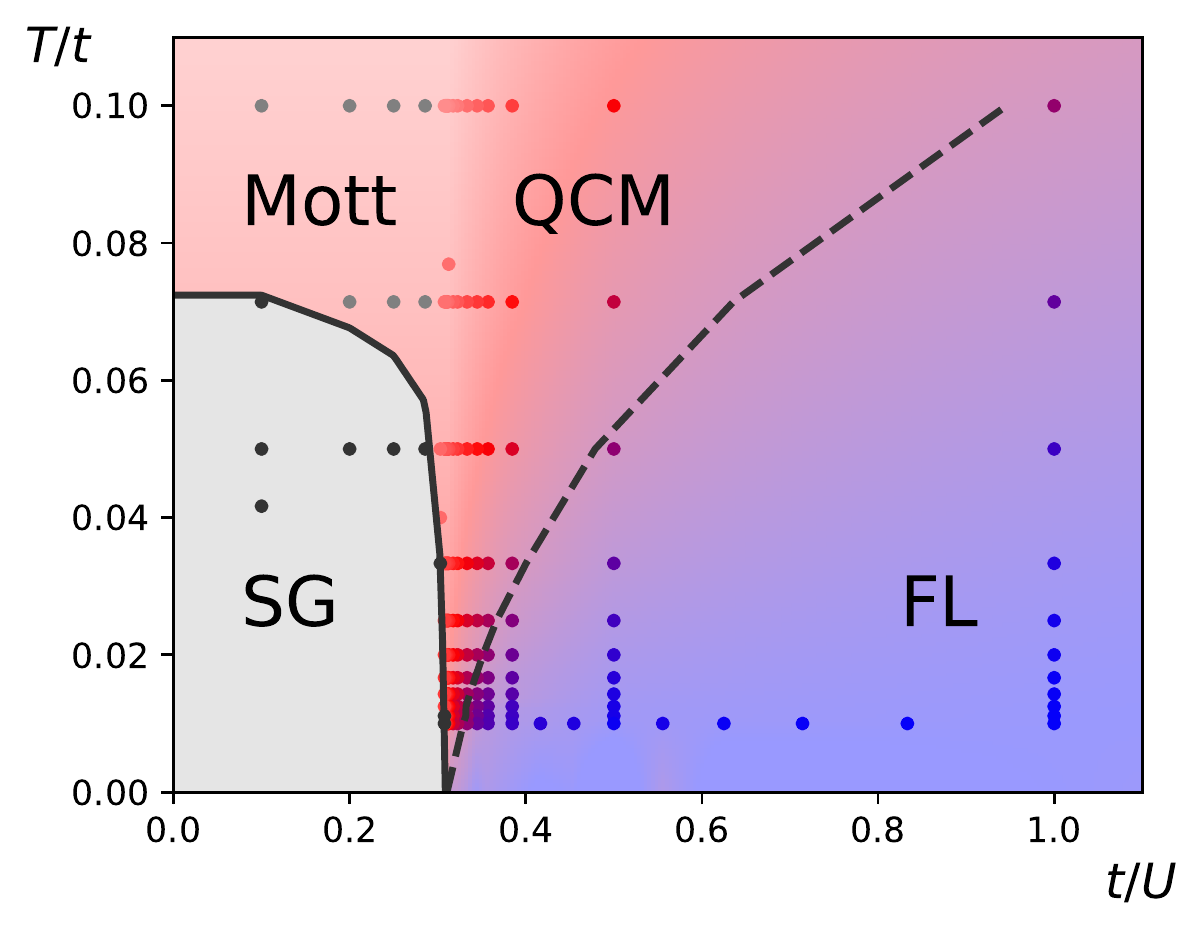}
\caption{
\label{fig:phdiag}
Calculated phase diagram of the $t$-$U$-$J$ model~(\ref{eq:model}]) at $J/t=0.5$.
Solid black curve indicates a 2nd-order phase transition to SG order.
Round markers represent parameters for which we have explicitly solved the model. Markers have been colored red where we find a Quantum Critical Metal (QCM) with QSL spin dynamics and blue where we find a FL. Background shading interpolates between the explicitly solved points. Dashed black curve indicates the crossover between QCM and FL regimes. Grey markers indicate Mott insulating solutions. Black markers indicate SG ordered solutions.
}
\end{figure}

We consider a `$t$-$U$-$J$ model' of itinerant spin-$1/2$ $SU(2)$ fermions with an on-site repulsive-$U$ Hubbard interaction and a random infinite-ranged spin-spin coupling, at half-filling.
Using the extended dynamical mean-field theory framework (EDMFT)~\cite{Sengupta95,smith_si_prb_2000,chitra_prl_2000,dmft2006} and numerical methods detailed below, 
we obtain the phase diagram displayed in Fig.~\ref{fig:phdiag}.   
At $t/U\ra0$, we have a Mott insulating SG phase (Fig.~\ref{fig:phdiag}), where the fermions are localized on-site and the model reduces to the disordered Heisenberg model. SG order is found below a freezing temperature $T_g\approx 0.14J$ for $t/U=0$ as previously established~\cite{bray1980,grempel1998} (See Supporting Information A and B).  
As $t/U$ is increased, the single-occupancy constraint is relaxed and the charge fluctuations lead to quantum-melting of SG order at a QCP $(t/U)_c\approx 0.31$ separating the SG from a FL phase at low enough temperature for $(t/U)>(t/U)_c$ (blue points in Fig.~\ref{fig:phdiag}).
Our key finding is a quantum critical region emanating from the QCP with QSL spin dynamics identical to that of the SY model~\cite{sy1993} and $T$-linear MFL scattering rate ${\rm Im}\Sigma(\omega\rightarrow0,T)\propto T$ (red points in Fig.~\ref{fig:phdiag}), leading to $T$-linear resistivity as shown below.

More precisely, our model Hamiltonian reads
\beq
H=-\sum_{\<ij\>,s=\uparrow,\downarrow}t_{ij}\,c^\dag_{is}c_{js} + U\sum_{i}n_{i\ua}n_{i\da} -\sum_{i<j}\frac{J_{ij}}{\sqrt{{\cal N}}}\vec{S}_i\cdot\vec{S}_j
\label{eq:model}
\eeq
In this expression, $J_{ij}$ are quenched random Heisenberg interactions~\cite{bray1980} drawn from a Gaussian distribution with $\<J_{ij}\>=0$ and $\<J_{ij}^{\ 2}\>=J^2$, ${\cal N}$ is the number of sites, and $\vec{S}_i=\tfrac{1}{2}c^\dag_{is}\vec{\sigma}_{ss'}c_{is'}$, with $\vec{\s}$ the Pauli matrices.
The model can either be formulated on the infinite-connectivity $z\rightarrow\infty$ Bethe lattice with $t_{ij}=t/\sqrt{z}$, or on a fully connected lattice with Gaussian distributed random $t_{ij}$'s with $\<t_{ij}\>=0$ and $\<t_{ij}^{\ 2}\>=t^2/{\cal N}$, leading to identical equations in the phase without magnetic ordering after replica averaging~\cite{dmft1996}. 
We restrict ourselves to the half-filling case $\m=U/2$ and choose $J=0.5t$. 
For a study of the $SU(M)$ version of this model in the large-$M$ limit at half-filling, see Ref.~\cite{Florens_2013}. 

To investigate the phase diagram of model (\ref{eq:model}), both the on-site repulsion in the charge-channel and the random interaction in the spin-channel need to be tamed. 
This is achieved using the EDMFT framework and the replica trick. In this framework, the calculation of the local Green's function and spin-spin correlation function is mapped onto the solution of a local `quantum impurity' problem subject to a self-consistency condition~\cite{bray1980,dmft1992,sy1993,dmft1996,Sengupta95,parcollet1999,GPS00,GPS01}.  
This mapping is exact in the infinite connectivity $z\rightarrow\infty$ or infinite volume limit ${\cal N}\rightarrow\infty$ of the two formulations of the model discussed above.

The resulting local effective action, after disorder averaging and making a replica diagonal ansatz, reads:
\begin{eqnarray}
\label{eq:LocalModel}
S_\text{eff}\,&=&\,-\,\beta\sum_{n,\s} c^\dag_\s \(i\w_n+\m-\Delta(i\w_n) \) c_\s\,+\,
U \int_0^\beta d\tau n_\ua n_\da \nonumber
\\
&-&\frac{J^2}{2}\int_0^\beta d\t d\t'\ Q(\t-\t')\vec{S}(\t)\cdot\vec{S}(\t').
\end{eqnarray}
In this expression, $\beta=1/T$ ($k_B=1$) is the inverse temperature, $\tau\in[0,\beta]$ stands for imaginary time and $\omega_n=(2n+1)\pi/\beta$ are Matsubara frequencies. 
The dynamical mean-field (hybridisation function) $\Delta$ and effective spin-spin retarded interaction $Q$ 
are subject to the following self-consistency conditions: 
\beq\label{eq:scc}
\Delta(\tau)\,=\,t^2 G(\tau)\,\,\,,\,\,\,
Q(\t-\t') &=\frac{1}{3}\,\left\langle\vec{S}(\t)\cdot\vec{S}(\t')\right\rangle
\eeq
in which the local Green's function $G(\tau) \equiv - \langle T c_\s (\tau) c^\dag_\s(0)\rangle$ 
and the local spin-spin correlator $\<\vec{S}(\t)\cdot\vec{S}(\t')\>$ are to be computed with the local 
effective action (\ref{eq:LocalModel}). 
Noting that $i \omega_n + \mu - \Delta(i\omega_n)$ is the inverse effective one-body propagator of this action, 
a fermionic self-energy can be defined from Dyson's equation as:
\beq
\Sigma(i\omega_n) = i \omega_n + \mu - \Delta(i\omega_n) - G^{-1}(i \omega_n).\label{eq:Dyson}
\eeq

 The local action (\ref{eq:LocalModel}) still presents a strongly correlated problem.
SY \cite{sy1993} made further progress on the random Heisenberg model by extending the spin symmetry to $SU(M)$ and taking the $M\rightarrow\infty$ limit, which allows for an analytical 
calculation of the spin-spin correlator of (\ref{eq:LocalModel}) and reduces the self-consistent problem to a non-linear integral 
equation. 
This was extended to itinerant fermions within the $t$-$J_{ij}$ model by Parcollet and Georges (PG) \cite{parcollet1999}, who obtained a FL regime of the doped model at low-$T$, and a quantum critical regime associated with the proximity of the spin-liquid Mott insulator characterized by a $\sqrt{\omega},\sqrt{T}$ self-energy but, remarkably, `bad metal' behaviour with linear resistivity.
Recently, fermionic versions of the random coupling problem, the so-called SYK models~\cite{kitaev1,sachdevPRX}, garnered much interest with again a solvable limit for a large number of flavors $M\rightarrow \infty$. 
Recent works~\cite{balents2017,syk-sc,pengfei_17,shenoy_18,McGreevy_18} extended the mechanism of PG \cite{parcollet1999} for linear-$T$ resistivity to a lattice of SYK `quantum dots' with hopping. 
Interestingly, when SYK dots are coupled to another band of otherwise free and translationally invariant (uniform hopping) fermions, not only does the $T$-linear resistivity extend down to zero temperature but the mechanism switches to that driven by the MFL $T$-linear scattering rate~\cite{senthil2018, patel2018}.

For the physical limit of a single flavor of spin-1/2 fermions that is of our interest,  the self-consistency equations above require computing two- and four-point correlators in the local model with $SU(2)$ symmetry.
We use an implementation~\cite{ctint} of Rubtsov's continuous-time interaction-expansion quantum Monte Carlo (CT-INT)~\cite{rubtsov2005} algorithm which is based on the TRIQS library~\cite{triqs}.
The algorithm works in imaginary time, so we will discuss most of our results directly on the imaginary axis without analytic continuation, except in the discussion of transport.
Our implementation determines the local spin-spin correlator  from the impurity three-point vertex function rather than through an operator insertion measurement.
This algorithmic improvement allows for a drastic speed-up of the calculations~\cite{ctint}.

\begin{figure}
\centering
\includegraphics[width=\linewidth]{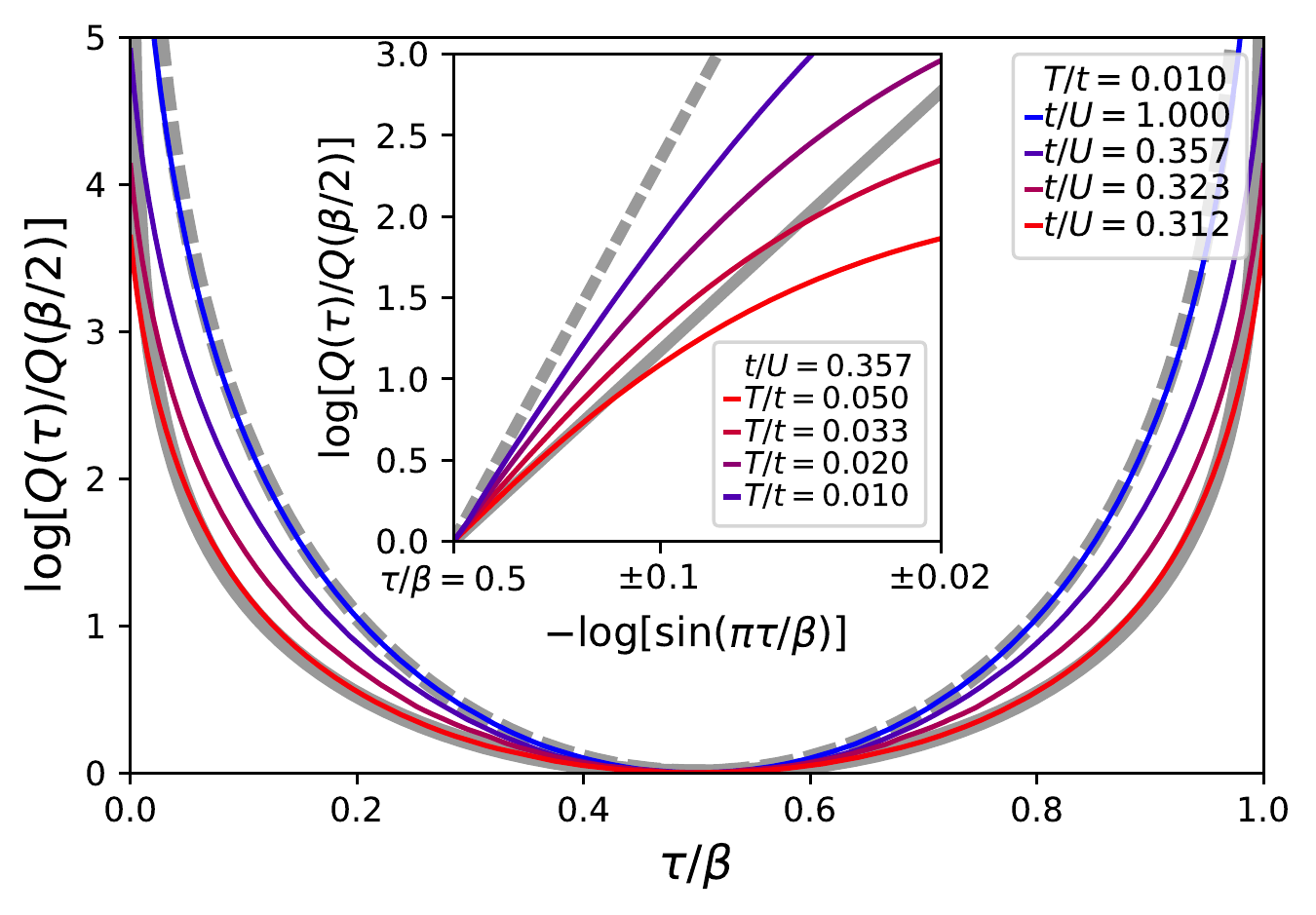}
\caption{
\label{fig:QSL}
Main: Spin susceptibility $\log[Q(\tau)/Q(\beta/2)]$ vs $\tau/\beta$ for $J/t=0.5$ and $T/t=0.01$, across several $t/U$.
Grey curves show $(1/\sin\pi\t/\b)^\alpha$ with $\alpha=1$ (solid) and $\alpha=2$ (dashed).
Color scheme follows the blue (FL) and red (QSL) gradient of Fig.~\ref{fig:phdiag}.
Inset:
Spin susceptibility $\log [Q(\tau)/Q(\beta/2)]$ vs $-\log[\sin(\pi\tau/\beta)]$, for $J/t=0.5$ and $t/U=0.357$, across a range of $T$,  demonstrating scaling behavior of $Q(\t)$ near $\tau=\b/2$.
Grey curves show $\alpha=1,2$ (solid, dashed).}
\end{figure}

Let us first consider the long time spin dynamics.
In Fig.~\ref{fig:QSL}, we display the local spin-spin correlation function $Q(\tau)$ at a fixed low temperature $T/t=0.01$, for various $t/U$ approaching the QCP at $(t/U)_c\approx 0.31$ from the FL limit cutting the phase diagram Fig.~\ref{fig:phdiag} along the horizontal axis.  
In the inset, we also display how $Q(\tau)$ varies upon raising temperatures for fixed $t/U=0.357$ making a vertical cut in the phase diagram slightly away from the QCP.
Since we work in the Matsubara formalism, a zero temperature long time asymptotic form $Q(t) \sim 1/t^\alpha$ transforms into a scaling function $Q(\tau) \sim \(1/\beta\sin(\pi\tau/\beta)\)^\alpha$ and the data should be examined near $\tau=\beta/2$.
Away from the critical point, for $t/U = 1.0$, we obtain the FL behaviour at long time $Q(t) \sim 1/t^2$ ($\alpha=2$).
The closer one gets to the critical point, the longer it takes to reach this asymptotic regime, reflecting the decrease of the FL coherence scale close to the critical point.
Once in the quantum critical regime, for $t/U = (t/U)_c \approx 0.31$, the long time spin dynamics crosses over to $Q(t) \sim 1/t$ ($\alpha=1$), which is the same power law as in the SY $M=\infty$ model. 
The QSL to FL crossover is also visible in the temperature cut shown in the inset, where we observe the crossover from
$1/t$ within the quantum critical fan above the Fermi liquid coherence temperature to $1/t^2$ at low-temperatures.
The phase classification at each point in Fig.~\ref{fig:phdiag} follows the above criterion to identify the FL regime and the QSL regime.

These results establish that our $SU(2)$ $t$-$U$-$J$ model has, in the quantum critical regime, the same QSL local spin dynamics ($\alpha=1$) as the SY model in the $M=\infty$ limit. 
Renormalisation group (RG) methods should prove useful in establishing 
analytically our numerical findings for $SU(2)$. For simplified versions of the effective action (\ref{eq:LocalModel}), e.g. involving only localized spins~\cite{Sengupta00}, RG methods have indeed established
\cite{Sengupta00,VBS2000,SBV1999,Vojta04,FritzVojta04,FritzFlorensVojta06,
FritzThesis,SS2001,VL2004,Si1993a,Si1993b}, 
that the $Q(t)\sim 1/t$ spin liquid behaviour 
is the only one consistent with the self-consistency condition (\ref{eq:scc}). 
This was recently extended to the QCP obtained by doping the $U=\infty$ model~\cite{DQCP2019}.

\begin{figure}
\centering
\includegraphics[width=.9\linewidth, trim=0cm 1cm 0cm 2cm, clip]{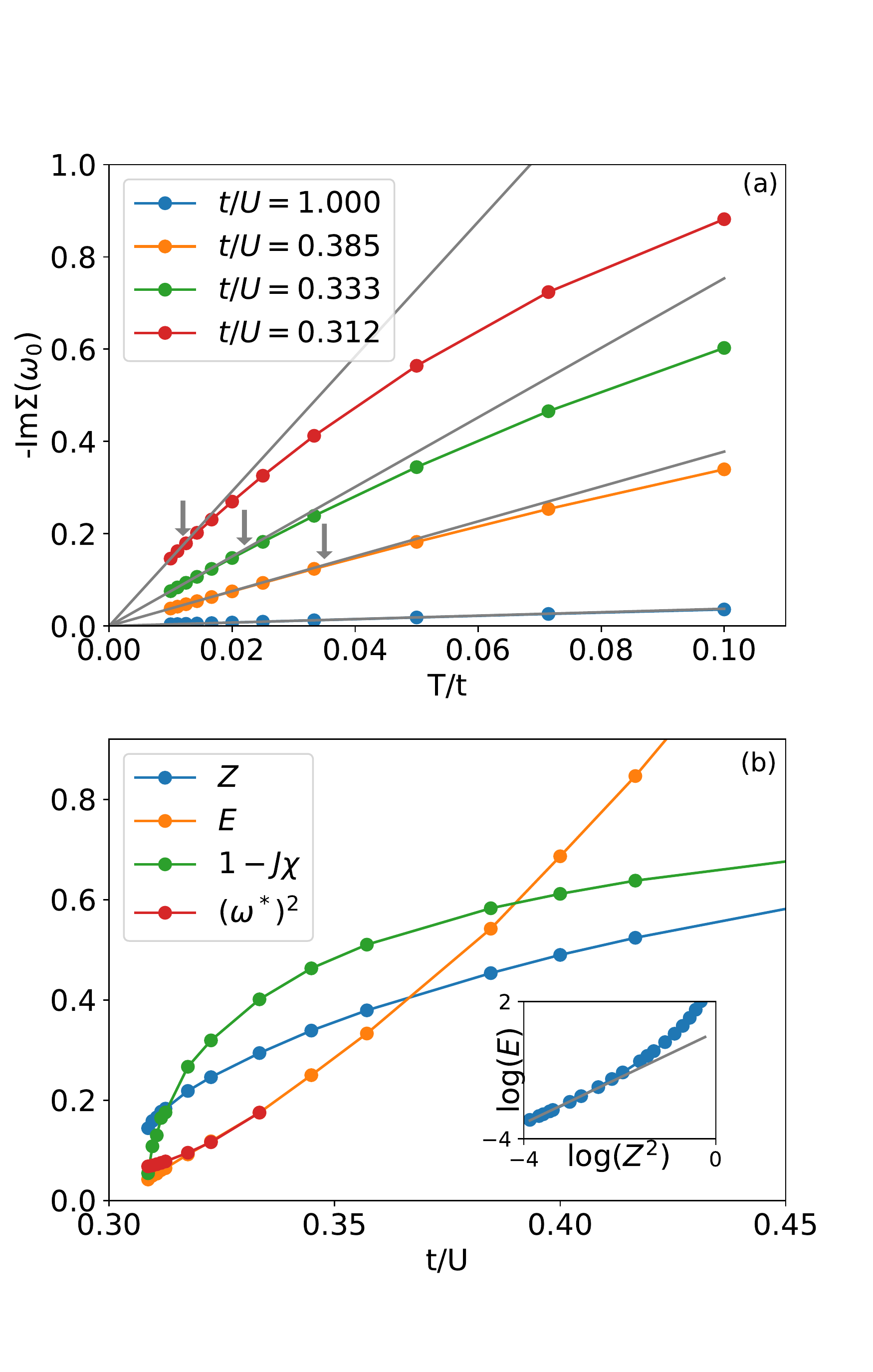}
\caption{\label{fig:FermiLiquid}
(a) Imaginary part of the self-energy at the first Matsubara point $-\im\Sigma(i\w_0=i\pi T)$ vs temperature $T$, for a range of $t/U$. Solid grey lines stand for the FL prediction of $\im\Sigma(i\w_0)\propto T$ from the lowest temperature. Arrows indicate the Fermi-liquid coherence temperature $T^*$ for each value of $t/U$. The solution at $t/U=1.0$ remains in the Fermi-liquid regime over the entire range of temperature considered. 
(b) Quasi-particle residue $Z$ and coherence scale $E$ as obtained by fitting (\ref{eq:SigmaFL}) to the self-energy data, ordering criterion for the SG phase $1-J\chi$, and the energy scale determined from scaling plot $(\omega^*)^2$, vs $t/U$.
Inset : $\log E$ vs $\log Z^2$ illustrating a dependency $E \propto Z^2$ close to the QCP. Grey line with slope 1 is plotted to guide the eye.
}
\end{figure}

\begin{figure}
\centering
\includegraphics[width=\linewidth, trim=2cm 1cm 2cm 2cm, clip]{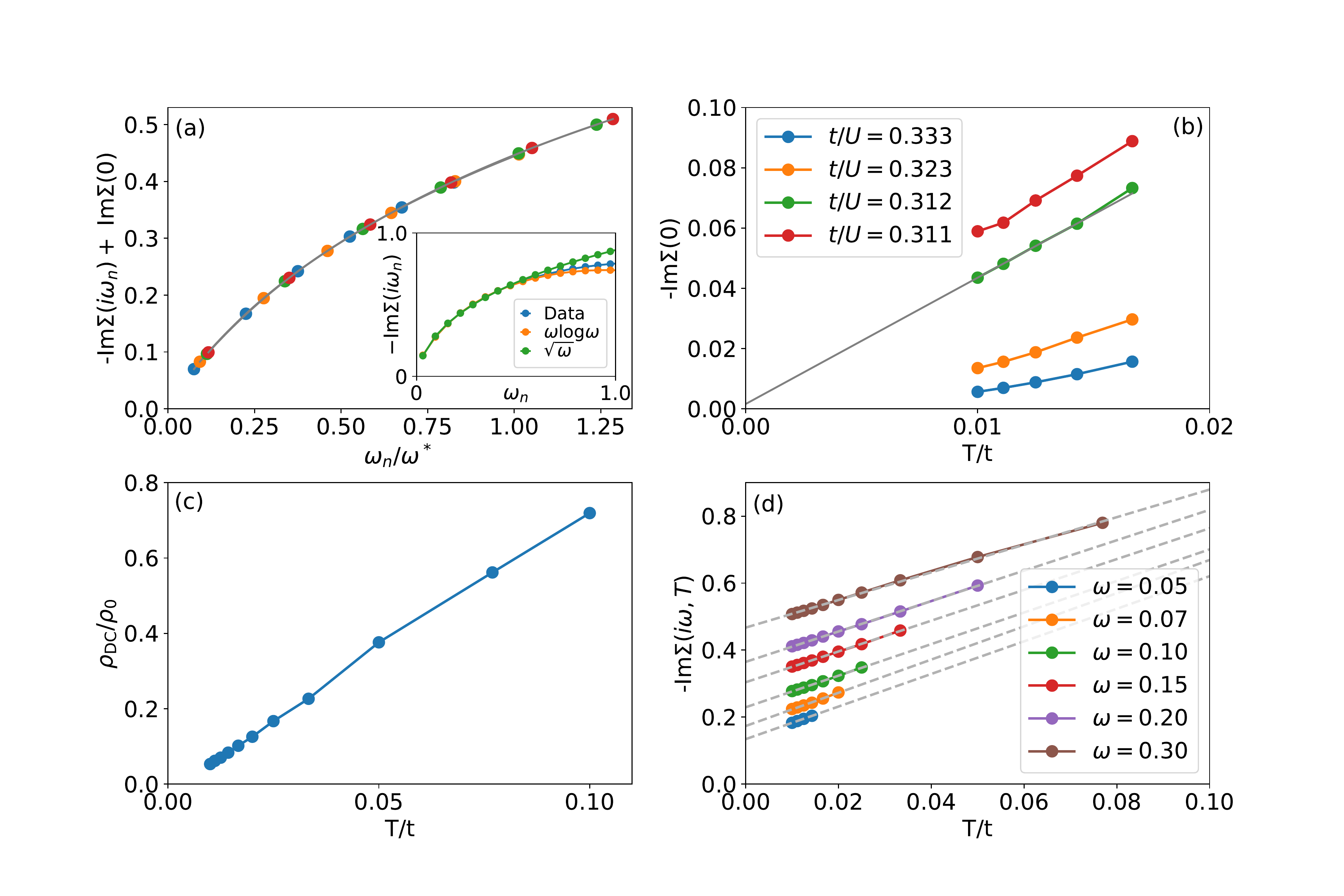}
\caption{\label{fig:scalingAndSig0}
(a) Imaginary part of self-energy with the scattering rate subtracted $-\(\im\Sigma(i\w_n)-\im\Sigma(0)\)$ vs the scaled frequency $\w/\w^*$ for various values of $U$ near the QCP at $T/t=0.01$, demonstrating the collapse onto the universal scaling function $f(\w/\w^*)$ (grey solid curve). Color scheme follows the legend of (b). Inset: Imaginary part of self-energy $-\im\Sigma(i\omega_n)$ vs Matsubara frequencies $\omega_n$ at the QCP $t/U=0.312$ and lowest accessible temperature $T/t=0.01$. Also shown are low-frequency fits of self-energy to the MFL form $c+a\omega_n\log\omega_n/b$ (orange) and the SYK form $c+a\sqrt{\omega_n}+b\omega_n$ (green).
(b) Scattering rate $-\im\Sigma(0)$ vs temperature $T/t$ at various values of $t/U$ near the QCP. At the QCP ($t/U=0.312$, green), the scattering rate is $T$-linear (linear fit in grey), in contrast to the quadratic behavior in the FL regime (blue).
(c) Resistivity $\rho_{DC}/\rho_0$ vs temperature $T/t$ at the QCP, computed with the analytically continued Green's function. The unit of resistivity is the MIR value $\rho_0=\hbar/e^2\phi(0)$, where $\phi$ is the transport function. 
(d) Imaginary part of self-energy at fixed, interpolated values of Matsubara frequency $-\im\Sigma(i\w=\text{fixed}, T)$ vs temperature $T/t$ at the QCP $t/U=0.312$, for various fixed values of frequency.
}
\end{figure}

Let us now consider the one particle properties, encoded by the self-energy $\Sigma$.
In the FL regime for $(t/U)_c\ll(t/U)$, the self-energy has the low energy expansion\footnote{In a FL, the real-frequency dependence of the self-energy is well-known to be $-\im\Sigma(\w)\sim (\w^2+(\pi T)^2)/E$. When the self-energy is analytically continued to Matsubara frequencies, the imaginary part of self-energy gains a linear term in frequency from the low-frequency expansion of $\re\Sigma(\w)$, so that $-\im\Sigma(i\w_n)\sim\w_n$.}
:
\begin{equation}\label{eq:SigmaFL} 
\im \Sigma(i \omega_n, T) \approx \left(1- \frac{1}{Z}\right) \omega_n + \frac{\omega_n^2-(\pi T)^2}{E} + O(\omega_n^3)
\end{equation}
In the small hopping limit $(t/U)\ll (t/U)_c$, $\Sigma$ diverges at low frequencies as $1/\omega_n$, indicating a transition into an insulating phase (see Supporting Information C). 
We examine the crossover from the FL to the quantum critical regime in several ways.
First, a direct consequence of (\ref{eq:SigmaFL}) is that the self-energy at the first Matsubara frequency is linear in temperature with vanishing quadratic corrections~\cite{chubukov_maslov_2012}: 
$\im\Sigma(i\w_0=i\pi T)=(1-1/Z)\pi T + \order(T^3)$. 
Deviation from linearity in $T$ at a temperature $T^*$ signals the FL coherence scale, and hence the crossover to the quantum critical regime.
This is illustrated on Fig. \ref{fig:FermiLiquid}a : when $t/U$ approaches $(t/U)_c$, the self-energy increases and $T^*$ (indicated by arrows on the figure) decreases.
More precisely, we extract the quasi-particle residue $Z$ and the coherence scale $E$ by fitting the functional form (\ref{eq:SigmaFL}) to the low-energy data using weighted least squares. 
Fig. \ref{fig:FermiLiquid}b shows that $Z$ and $E$ vanishes at the QCP. 
The susceptibility to SG order is given by~\cite{GPS00,GPS01} $\chi_{sg}\propto \chi^2/(1-J^2\chi^2)$ with $\chi$ the local susceptibility. 
As shown in Fig.~\ref{fig:FermiLiquid}, we find that $1-J\chi$ also vanishes close to the QCP, 
indicating the boundary of the SG phase. 
Within our numerical accuracy, we cannot however exclude that $1-J\chi$ vanishes at a slightly larger value of $t/U$ than $E$, possibly indicating a small intervening region of {\it metallic} SG~\cite{DQCP2019}.

In order to analyse the quantum critical point, we attempt to scale the self-energy for $t/U$ close to $(t/U)_c$, for our lowest temperature $T/t=0.01$, with an ansatz of the form 
\begin{equation}\label{eq:ScalingForm}
\im \Sigma(i \omega_n) \approx \im \Sigma(0) + f\left( \dfrac{\omega_n}{\omega^*} \right)
\end{equation}
which applies for $\omega_n$ and $\omega^*$ smaller than the high-energy cutoff $J$, but $\omega_n/\omega^*$ otherwise arbitrary.
We determine numerically $\im \Sigma(0)$, $\omega^*$ and the scaling function $f$ by requesting that optimal data collapse is obtained, using a minimisation procedure.
We obtain a remarkable collapse of the data, presented in Fig. \ref{fig:scalingAndSig0}a, with $\omega^*$  presented in Fig. \ref{fig:FermiLiquid}b.

For  $\omega < \omega^*$, the ansatz (\ref{eq:ScalingForm}) has to reproduce (\ref{eq:SigmaFL}), which implies $Z\propto \omega^*$ (for $\omega^* \rightarrow 0$), and $E \propto (\omega^*)^2$, hence $E \propto Z^2$, as illustrated in the insert of \ref{fig:FermiLiquid}b.
Note however that the $\omega^*$ obtained from the data collapse does not perfectly vanish close to the QCP, which may be due to numerical uncertainty, or possibly to a  weakly first order transition or to an intervening metallic SG phase 
as mentioned above. 
In the quantum critical regime, i.e. for $\omega > \omega^*$, the self-energy is very well described by a MFL form $\im\Sigma(\w_n) \propto \Sigma(0) + a\omega_n\ln\omega_n/b$. (See inset of Fig.~\ref{fig:scalingAndSig0}a.)
However, the low temperature behaviour obtained in the large-$M$ limit \cite{sy1993}, i.e. $\sqrt{\omega_n}$, cannot be excluded given our data.
Indeed, the CT-INT algorithm is faced with a sign problem at low-$T$ which prevents us from reaching the very low temperature regime required to settle this question.
This conclusion holds both for the scaling function $f$, and for a direct analysis of the self-energy at $t/U=(t/U)_c$.

The value of the self-energy at zero frequency $\im \Sigma(0)$ is of crucial importance for transport properties.
In Fig. \ref{fig:scalingAndSig0}c, we show $\im \Sigma(0)$ extracted from the scaling analysis, for various $U$ close to the QCP.
We find that $\im \Sigma(0) \propto T$ at low temperature at the QCP.
This is confirmed in Fig. \ref{fig:scalingAndSig0}d:
$\im \Sigma(i\omega)$ obtained (by interpolation) for {\it fixed}
imaginary frequency $i \omega$ is linear with temperature, with a slope weakly dependent on the frequency.

Let us finally turn to the DC resistivity in the quantum critical region.
The Kubo formula reduces to the polarization bubble 
(vertex corrections vanish in this quantity in DMFT)
\begin{align}%\label{eq:resistivity}
\sigma_{\rm DC}=
\frac{2\pi e^2}{\hbar}\int d\omega \frac{\beta}{4\cosh^2(\b\w/2)}\,
\int d\e\, \ph(\e) A(\e,\w)^2
\end{align}
In this expression, $\e$ is the energy of a bare single-particle state within the band, $A(\e,\w)=-(1/\pi)\im G^R(\e, \w)$ 
is the energy (momentum-) resolved spectral function and 
$\phi(\e)$ is the transport function 
$\phi(\e)=\sum\limits_{\mathbf{k}}\(\partial\e_{\mathbf{k}}/\partial k_x\)^2\d(\e-\e_{\mathbf{k}})$, which we take to be the sum-rule preserving expression on the Bethe lattice (see e.g.~\cite{deng_badmetals_prl_2013}): $\phi(\e)=\phi(0)[1-\(\e/2t\)^2]^{3/2}$. 
To obtain $\sigma_{\rm DC}$, we perform an analytic continuation of the Monte Carlo data using Pad\'e approximants~\cite{vidberg1977} to obtain the real-frequency self-energy $\Sigma(\w)=\Sigma'(\w)+i\Sigma''(\w)$ and the spectral function: 
$\pi A(\e,\w)=-\Sigma''(\w)/[(\w+\mu-\e-\Sigma'(\w))^2+\Sigma''(\w)^2]$. 
The resulting resistivity $\rho_{\rm DC}=1/\sigma_{\rm DC}$ vs temperature $T$ is plotted in Fig.~\ref{fig:scalingAndSig0}c, clearly consistent with $T$-linear resistivity within numerical accuracy.

The origin of this behaviour can be directly related to the $T$-linear behaviour of the scattering rate $\Sigma''(0)$. 
Indeed, observing that the latter is a much smaller scale than the bandwidth at low-$T$, the 
integral over $\e$ can be approximated as:
\beq\label{eq:drude1}
\int d\e\, \ph(\e) A(\e,\w)^2\,\sim\,\frac{\phi\left[\w+\mu-\Sigma'(\w)\right]}{2\pi|\Sigma''(\w)|}
\eeq
Due to the Fermi factor only $|\w|\lesssim T$ is relevant for the frequency 
integral, so that the right hand side of this expression can be replaced by its Fermi surface 
contribution $\omega=0$ (see Supporting Information D). 
Observing that $\mu-\Sigma'(0)=0$, we finally obtain:
\beq
\sigma_{\rm DC}\,=\,\frac{e^2\phi(0)}{\hbar}\int \frac{\b d\w}{4\cosh^2(\b\w/2)}\frac{1}{\lb\Sigma''(\w)\rb}\sim \frac{e^2\phi(0)}{\hbar T}.\label{eq:drude2}
\eeq
$\rho_0=(\hbar/e^2)/(\phi(0)/t)$ can be taken as the order of magnitude of the MIR resistivity~\cite{deng_badmetals_prl_2013}, 
so that we obtain at the QCP $\rho_{\rm DC}/\rho_0 \sim T/t$ down to the lowest value of $T$ we could reach.

We would like to emphasize that both the mechanism and the physical meaning of this 
$T$-linear resistivity are different from the ones reported in Ref.~\cite{parcollet1999} and in 
the SYK $M\ra\infty$ lattice models~\cite{balents2017,syk-sc,pengfei_17,shenoy_18,McGreevy_18}. 
There, the scattering rate had a $\sim \sqrt{T}$ temperature dependence and dominated the band dispersion 
in the incoherent metal regime $T>T^*$, resulting in the resistivity being proportional by the square of 
the scattering rate and larger than the MIR value. 
Here in contrast, the scattering rate is $T$-linear (Planckian) and small at low $T$, and the 
band dispersion dominates, resulting in linear resistivity down to low $T$. 
The present mechanism is also distinct from the generic bad metal behaviour of lattice models 
at very high $T$ comparable to the bandwidth~\cite{palsson_prl_1998,gunnarsson_rmp_2003,deng_badmetals_prl_2013,
perepelitsky_hiT_transport_prb_2016,cha2019}: there, the scattering is constant and the $T$-linear behaviour 
is associated with the $T$-dependence of thermodynamic quantities such as the kinetic energy $\sim 1/T$ which 
play the role of an effective carrier number. We have checked (see Supporting Information E) that in contrast the kinetic energy of our model is constant in the range of $T$ of interest. 

In this work, we considered the insulator to metal transition and quantum-melting by charge fluctuations of the spin-glass ground-state of the $SU(2)$ random-bond Heisenberg model. 
At the QCP separating the spin-glass from the Fermi liquid, we find 
a non-Fermi liquid state  
with long-lived spin correlations $\<\vec{S}(t)\cdot\vec{S}(0)\>\sim 1/t$, 
(as in the large-$M$ limit of the $SU(M)$ SY model)
and a  $T$-linear resistivity arising from a $T$-linear (Planckian)
scattering rate $\im\Sigma(\w=0,T)\propto T$.
In the temperature range accessible in this work, this quantum critical regime is compatible with a marginal Fermi phenomenology $\Sigma(\w)\sim -\w\log \w$. 
Fully establishing this behaviour down to 
zero temperature may require a new generation of quantum impurity solvers 
such as real time diagrammatic Monte Carlo~\cite{Profumo_2015, Bertrand_PRX2019}. 
Another open question is whether our results for the scattering rate 
also apply to the doped case recently considered in Ref.~\cite{DQCP2019}. Also finding the renormalization group fixed point associated with our metal insulator transition quantum critical point remains an open question.

\vspace{1ex}
\noindent{\bf Acknowledgements:} We are grateful to Chao-Ming Jian, Aavishkar Patel, Subir Sachdev, and Anirvan Sengupta for useful discussions. E-AK and PC are supported by the U.S. Department of Energy, Office of Basic Energy Sciences, Division of Materials Science and Engineering under Award DE-SC0018946.
The Flatiron Institute is a division of the Simons Foundation.

\bibliographystyle{apsrev4-1}
\bibliography{bib}

\clearpage

\appendix 

\section{Supporting Information}

The following sections contain supporting plots and information to the discussion in the main text.

\section*{A. Benchmarking our methods in the $t/U\rightarrow0$ limit}
\begin{figure}
\centering
\includegraphics[width=0.5\textwidth]{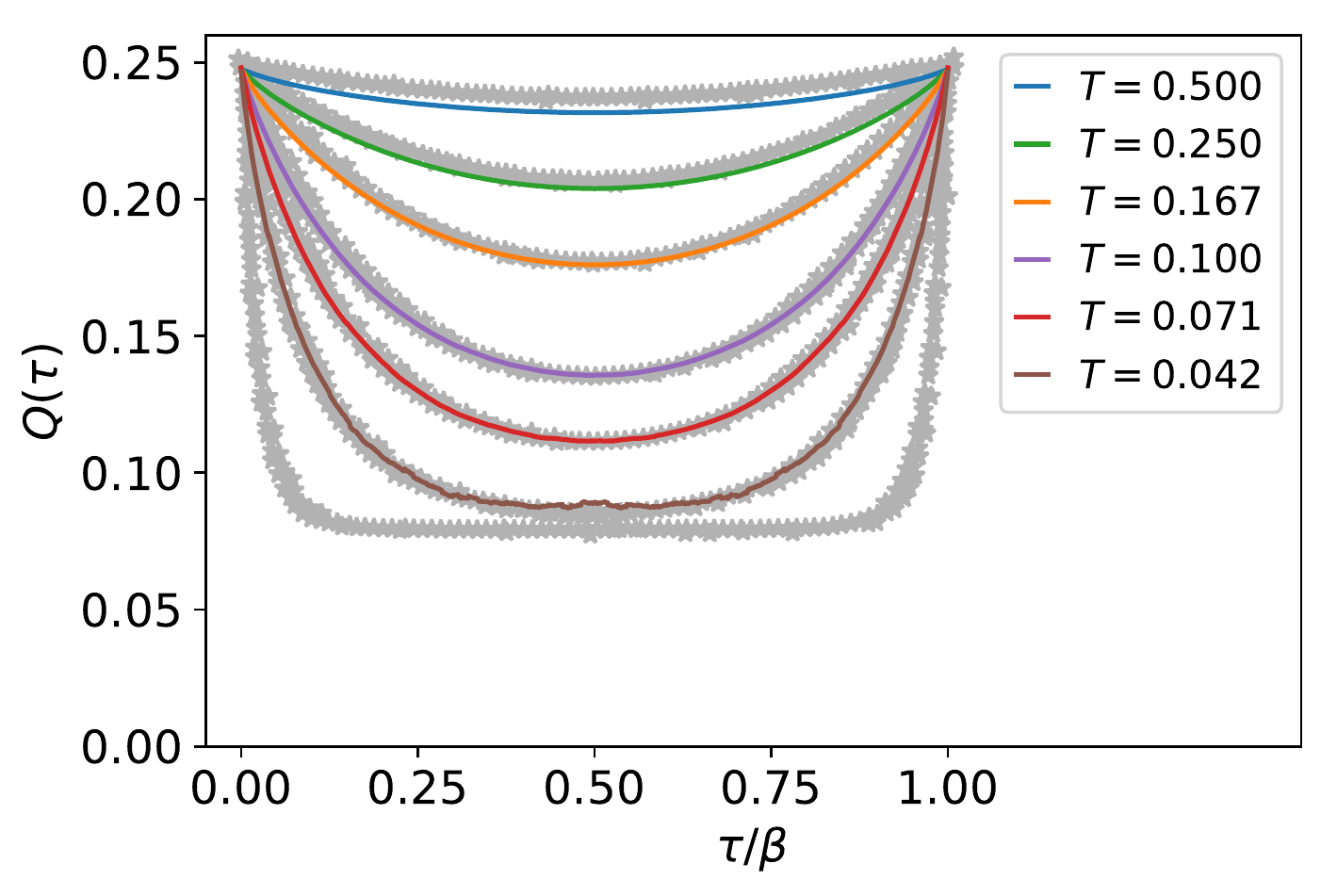}
\caption{
\label{fig:GRValidation}
Plot of the spin-spin correlator $Q(\tau)$ vs Matsubara time $\t/\beta$ at $t/U=0.1$ and various temperatures, laid over the result from~\cite{grempel1998} in grey.
}
\end{figure}

In Fig.~\ref{fig:GRValidation}, we plot the spin-spin correlator $Q(\t)$ deep in the Mott insulator regime $t/U=0.1$, where the fermions are localized on-site (in the paramagnetic phase). We show data from the same temperatures presented in~\cite{grempel1998} and overlay our results on their plot, shown in grey. We are not able to reproduce Grempel \& Rozenberg's results at lowest temperatures due to a sign problem in CT-INT.

We find the results to be in excellent quantitative agreement. The small discrepancy between our results and Grempel \& Rozenberg's at high temperatures $T=0.5$ is due to the single-occupancy constraint not being perfectly realized at finite itinerancy $t/U>0$, and we observe the discrepancy at fixed temperature vanishing as $t/U$ is decreased (analysis not shown).

\section*{B. SG phase transition}
\begin{figure}
\centering
   \includegraphics[width=0.5\textwidth]{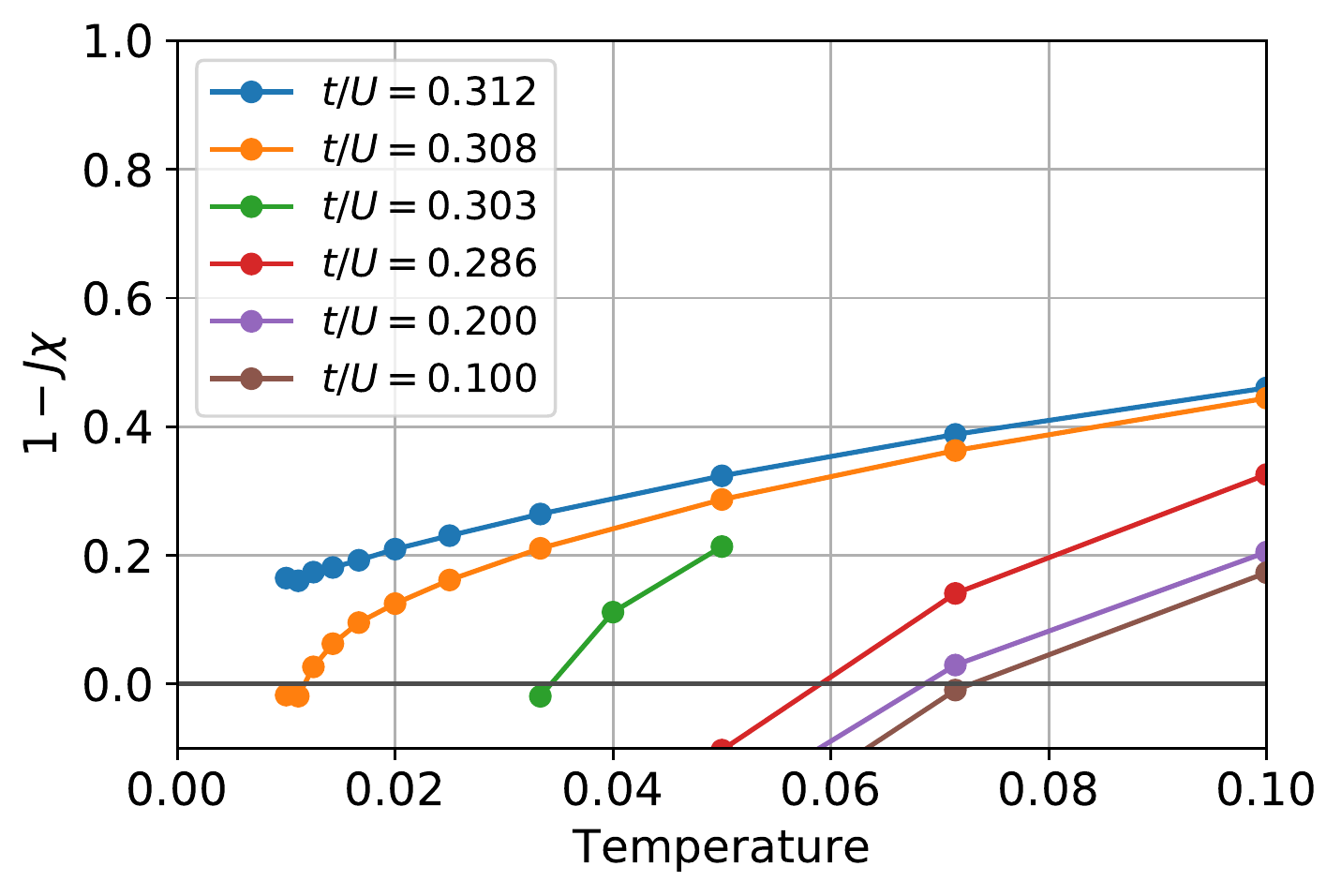}
\caption{
   \label{fig:jchi} $1-J\chi_{loc}$ vs temperature.
   The spin glass phase transition  corresponds to  $1- J\chi_{loc}=0$. It is present at values of $t/U$ less than the critical interaction strength $(t/U)_c=0.312$.
}
\end{figure}

The criterion for transition from a paramagnet to the spin glass  phase is  $J\chi_{loc}=1$~\cite{bray1980}, where $\chi_{loc}=\int_0^\beta d\t\ Q(\t)$ is the local spin susceptibility. We plot $1-J\chi$ vs temperature for various values of $t/U$ in Fig.~\ref{fig:jchi} (so that the spin glass phase  is determined by $1-J\chi<0$) showing how the SG transition temperature $T_g$ shifts with $t/U$.

In the non-itinerant limit $t/U\ra0$, where the fermions are localized on-site and the model reduces to the disordered Heisenberg model, we recover the results of Grempel \& Rozenberg $T_g\approx0.071$ ~\cite{grempel1998}. As $t/U$ is decreased and the localization constraint is relaxed, the spin glass transition temperature decreases to 0 at the QCP.

\section*{C. Crossover to the Mott insulator}
\begin{figure}
\centering
   \includegraphics[width=0.5\textwidth]{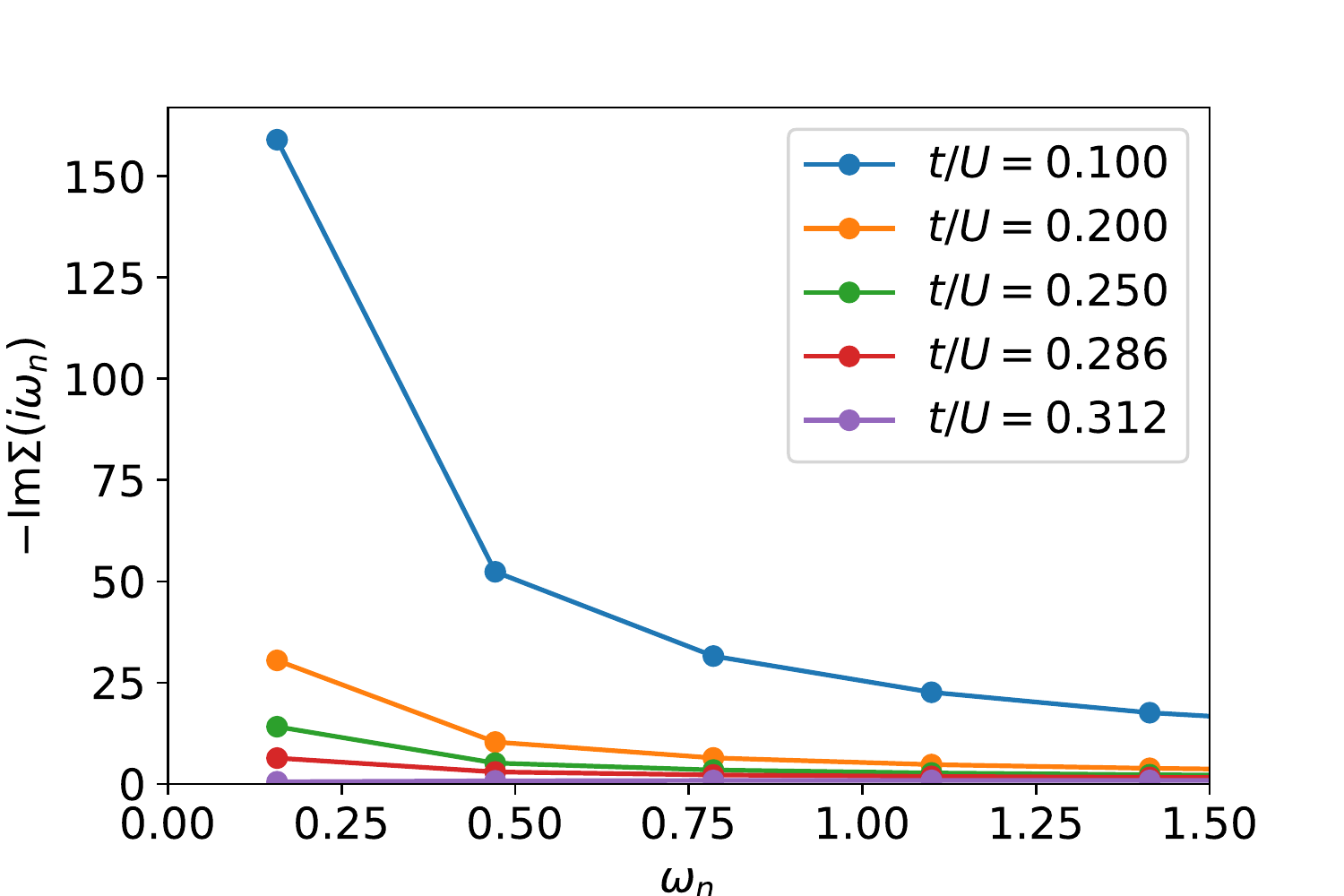}
\caption{
\label{fig:insulator}
Self-energy $-\im\Sigma(i\w_n)$ vs $\w_n$ at $T=0.05$ for a range of $t/U$, showing the crossover to a Mott insulator, when the self-energy diverges at low-frequency $-\im\Sigma\approx U^2/4i\w_n$.
}
\end{figure}

Across the critical interaction $(t/U)_c\approx 0.312$, the low-energy behavior of the self-energy $-\im\Sigma(i\w_n)$ crosses over from decreasing with decreasing frequency to increasing. This crossover is demonstrated at $T=0.05$ in Fig.~\ref{fig:insulator}. At the QCP $t/U=0.312$, the self-energy at low-frequency is on the order of temperature $T$. By contrast, at smaller values $t/U$, the self-energy is diverging at low-frequency and approaches the Mott insulator form $\Sigma(i\w_n)\approx U^2/(4i\w_n)$.

\section*{D. Analysis of Kubo formula}
\begin{figure}
\centering
\includegraphics[width=0.5\textwidth]{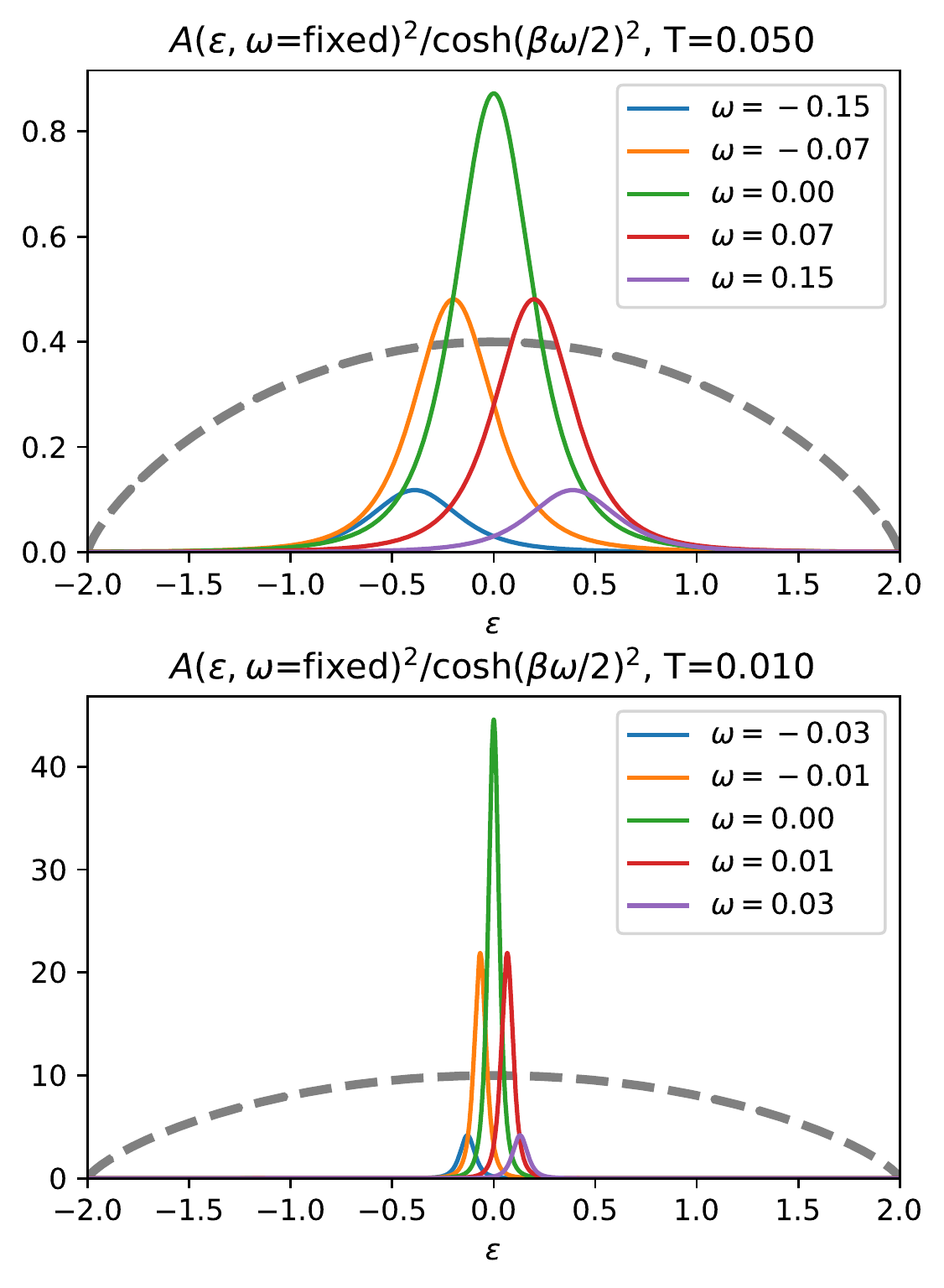}
\caption{\label{fig:infbw}
The momentum integrand of the Kubo formula $A(\e,\w)/\cosh(\b\w/2)^2$ at high ($T=0.05$) and low ($T=0.01$) temperatures and various frequencies $|\w|\leq 3T$ at the QCP $t/U=0.312$. Dashed grey shows the transport function $\phi(\e)$, up to a scale factor. 
}
\end{figure}

In this section, we give a detailed derivation of $T$-linear resistivity from the Kubo formula, expanding upon the discussion in the main text.
\beq
\sigma_{\rm DC}=&
\frac{2\pi e^2}{\hbar}\int d\omega \frac{\beta}{4\cosh^2(\b\w/2)}\,
\int d\e\, \ph(\e) A(\e,\w)^2\\
=&
\frac{2\pi e^2}{\hbar}\int d\omega \frac{\beta}{4\cosh^2(\b\w/2)}\nonumber\\
&\int d\e\, \phi(\e)\(\frac{1}{\pi}\frac{\Sigma''(\w)}{(\w+\mu-\e-\Sigma'(\w))^2+\Sigma''(\w)^2}\)^2
\eeq
In Fig.~\ref{fig:infbw} we have plotted the Kubo formula integrand $A(\e,\w)^2/\cosh(\b\w/2)^2$ vs $\e$ at various fixed frequencies $\omega$, at high-temperature (upper, $T=0.05t$) and low-temperature (lower, $T=0.01t$). We have also plotted in dashed grey the transport function $\phi(\e)$ up to a scale factor for clarity. Note that $A(\e, \w)$ is simply a Lorentzian in $\e$ of width $\Sigma''(\w)$ centered at $\w+\m-\Sigma'(\w)$.

We observe that $\phi(\e)$ is much wider than $A(\e,\w)^2$, so that $\phi(\e)\approx \phi(0)$ whenever $A(\e, \w)$ is appreciably non-zero. This is because the width $\lb\Sigma''(\w)\rb$ and center location $\lb\w+\m-\Sigma'(\w)\rb$ of the Lorentzian spectral function are both of order $T$ and thus much smaller than the bandwidth $4t$. Thus we make the replacement $\phi(\e)\ra\phi(0)$ in the Kubo formula.
\beq
\sigma_{\rm DC}=&
\frac{2\pi e^2\phi(0)}{\hbar}\int d\omega \frac{\beta}{4\cosh^2(\b\w/2)}\nonumber\\
&\int d\e\, \(\frac{1}{\pi}\frac{\Sigma''(\w)}{(\e-(\w+\mu-\Sigma'(\w)))^2+\Sigma''(\w)^2}\)^2\\
=&
\frac{2\pi e^2\phi(0)}{\hbar}\int d\omega \frac{\beta}{4\cosh^2(\b\w/2)}\nonumber\\
&\int d\e\, \(\frac{1}{\pi}\frac{\Sigma''(\w)}{\e^2+\Sigma''(\w)^2}\)^2\\
=&
\frac{2\pi e^2\phi(0)}{\hbar}\int d\omega \frac{\beta}{4\cosh^2(\b\w/2)}\(\frac{1}{2\pi}\frac{1}{\Sigma''(\w)}\)
\eeq
The Fermi factor $1/4\cosh^2(\b\w/2)$ cuts off the frequency integral to $|\w|\lesssim T$. In this restricted frequency window, the imaginary part of self-energy $\Sigma''(\w)$ is close to its zero-frequency value $\Sigma''(0)\sim T$.
\beq
\sigma_{\rm DC}=&
\frac{e^2\phi(0)}{\hbar}\int d\omega \frac{\beta}{4\cosh^2(\b\w/2)}\frac{1}{\Sigma''(\w)}\\
\sim&
\frac{e^2\phi(0)}{\hbar}\frac{1}{\Sigma''(0)}\int d\omega \frac{\beta}{4\cosh^2(\b\w/2)}\\
\sim&
\frac{e^2\phi(0)}{\hbar T}
\eeq

\section*{E. Kinetic energy}

\begin{figure}
\centering
\includegraphics[width=0.5\textwidth]{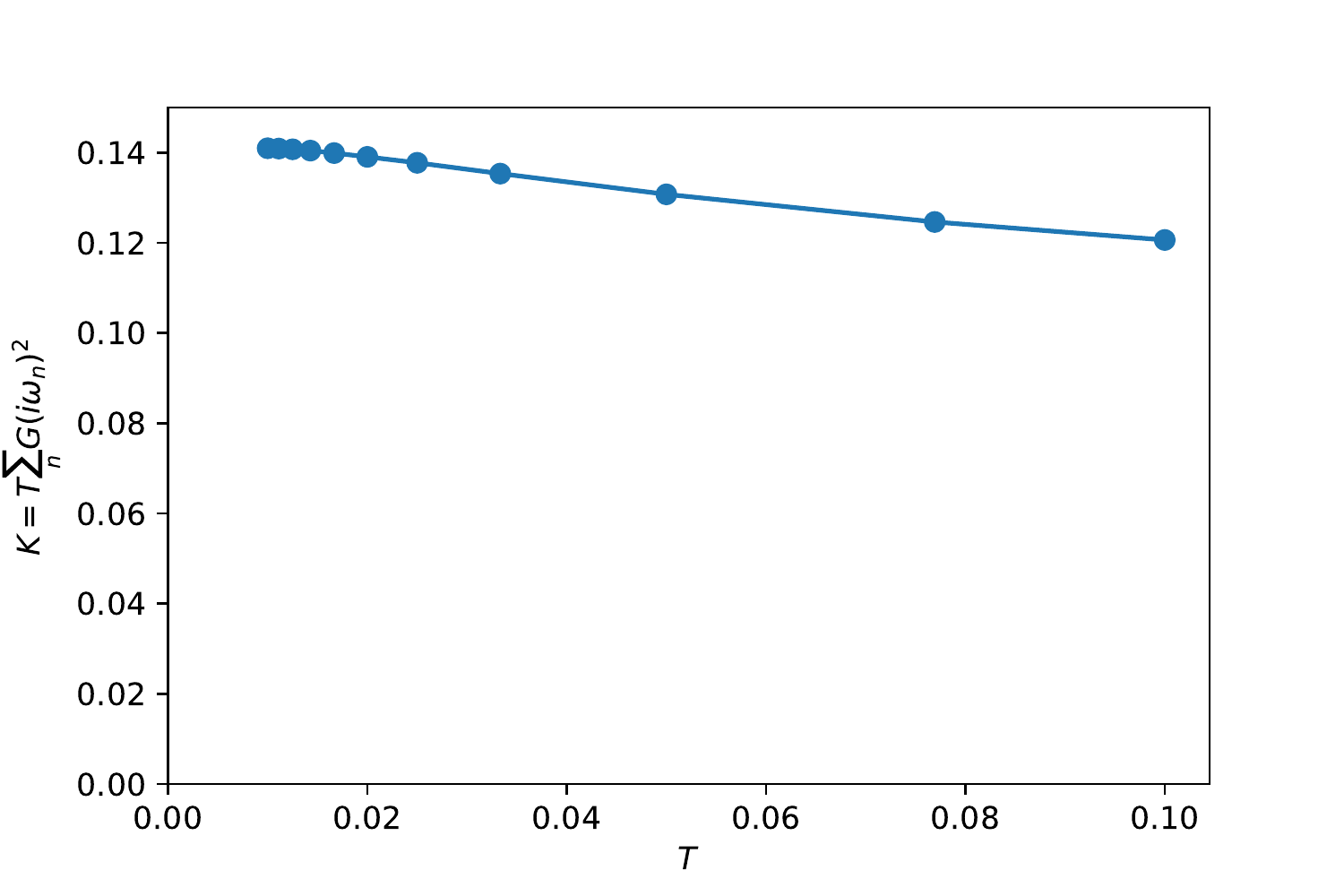}
\caption{\label{fig:kinetic}
The kinetic energy of the electrons on the Bethe lattice $K=T\sum\limits_{n}G(i\omega_n)^2$ vs temperature T at the QCP $t/U=0.312$.
}
\end{figure}

The DC conductivity can be written in the form $\sigma_{DC}=\mathcal{D}\tau$, where $\mathcal{D}$ is the Drude weight or carrier number and $\tau$ is the transport lifetime. The Drude weight is defined by the integral of optical conductivity near zero-frequency. If transport processes with large energy transitions are suppressed, the optical conductivity at high-frequency can be neglected and the Drude weight is then well-approximated by the kinetic energy $\mathcal{D}\approx K=\int_{-\infty}^{\infty}\sigma(\omega)d\omega$.

On the Bethe lattice, the kinetic energy can be further reduced to the expression $K=T\sum\limits_{n}G(i\omega_n)^2$, which we plot in Fig.~\ref{fig:kinetic}. We find that the kinetic energy at the QCP is nearly $T$-independent in the temperature range we consider. As the conductivity in the same parameter regime is $\sim1/T$, we conclude that the transport lifetime in our MFL is Planckian $\tau\sim \hbar/k_BT$.

\section*{F. Self-energy at the QCP}

\begin{figure}
\centering
\includegraphics[width=0.5\textwidth]{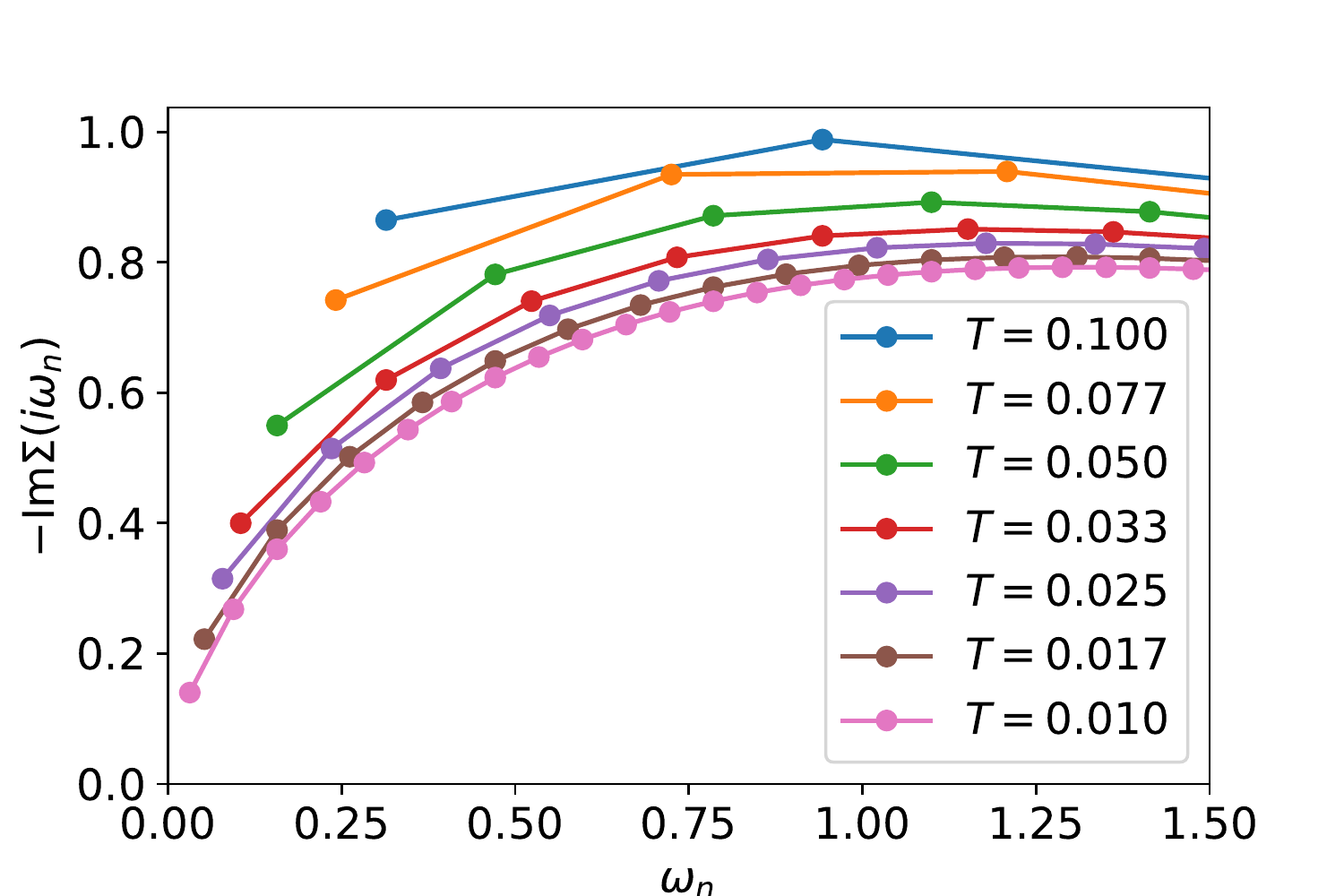}
\caption{\label{fig:mats-sig} Imaginary part of self-energy $-\im\Sigma(i\w_n)$ vs Matsubara frequencies $\w_n$ for a range of temperatures at the QCP $t/U=0.312$.}
\end{figure}

\begin{figure}
\centering
\includegraphics[width=0.5\textwidth]{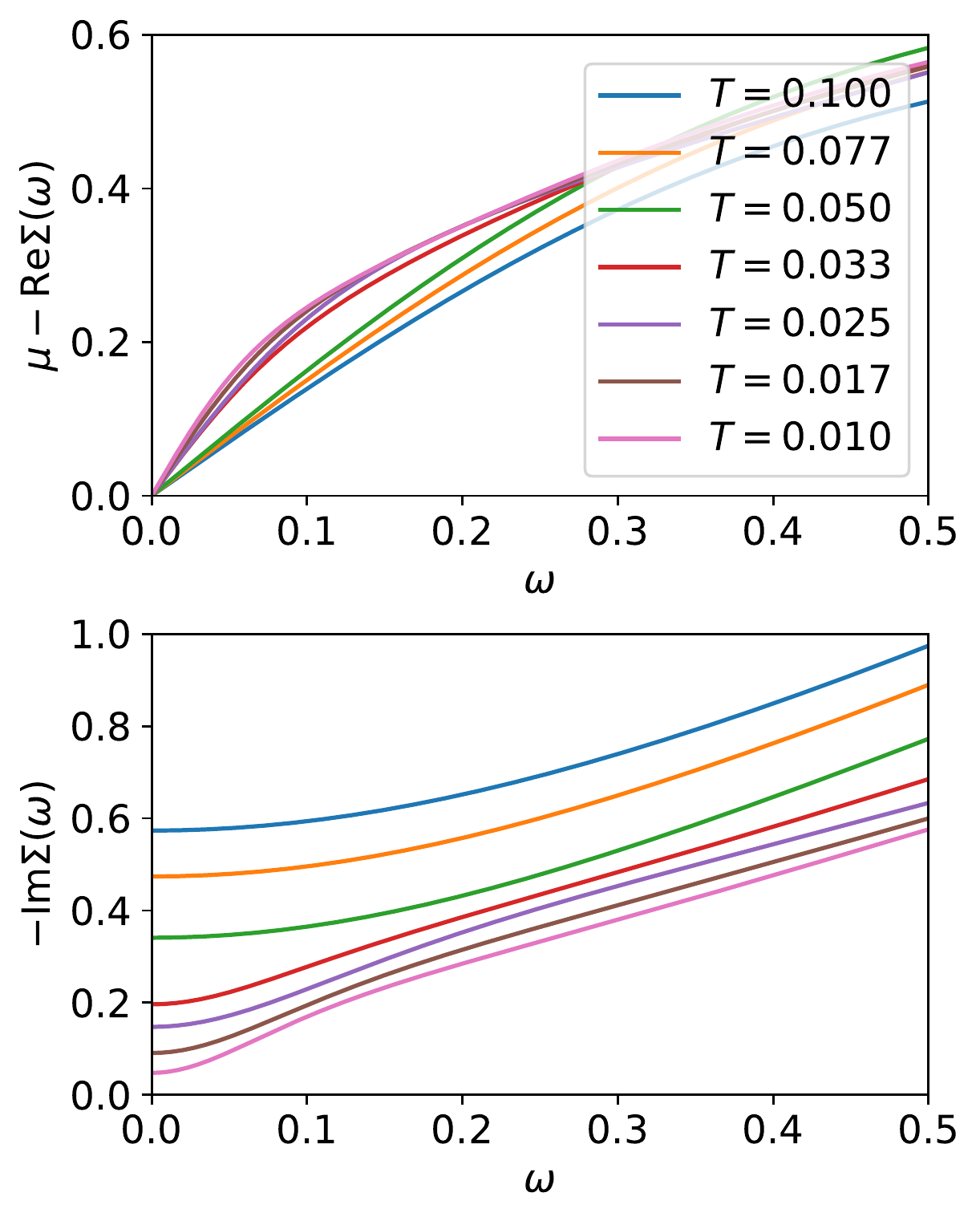}
\caption{\label{fig:real-sig} Pad\'e analytically continued self-energy, real $\m-\re\Sigma(\w)$ (upper) and imaginary $-\im\Sigma(\w)$ (lower) vs real frequencies $\w$ for a range of temperatures at the QCP $t/U=0.312$.}
\end{figure}

In Figs.~\ref{fig:mats-sig},\ref{fig:real-sig}, we plot the self-energy at the QCP for various temperatures. Fig.~\ref{fig:mats-sig} shows the Matsubara frequency self-energy from CT-INT $-\im\Sigma(i\w_n)$ vs $\w_n$, whereas Fig.~\ref{fig:real-sig} shows the real-frequency self-energy from Pad\'e analytic continuation $\mu-\re\Sigma(\w)$, $-\im\Sigma(\w)$ vs $\w$.

\clearpage
\end{document}